# Synthesis and Hydrogen Sorption Characteristics of Mechanically Alloyed Mg(Ni$_x$Mn$_{1-x}$)$_2$ Intermetallics

Evangelos I Gkanas[1*], Martin Khzouz[1], Alina Donac[2], Alexandra Ioannidou[3], George Stoian[2], Nicoleta Lupu[2], Margaritis Gjoka[3], Sofoklis S Makridis[4]

[1] *Hydrogen and Mobility Lab, Centre for Mobility and Transport, Coventry University, Coventry University, Priory Street, Coventry, CV1 5FB, United Kingdom*
[2] *National Institute of Research and Development for Technical Physics, Iasi, Romania*
[3] *Institute for Advanced Materials, Physicochemical Processes, Nanotechnology & Microsystems, National Centre for Scientific Research 'Demokritos', Agia Paraskevi, Athens, 15310, Greece*
[4] *Department of Environmental and Natural Resources Management, University of Patras, 2 Seferi St. Agrinio, Greece*

*Email: evangelos.gkanas@coventry.ac.uk*

**Abstract**

New ternary Mg-Ni-Mn intermetallics have been successfully synthesized by High Energy Ball Milling (HEBM) and have been studied as possible materials for efficient hydrogen storage applications. The microstructures of the as-cast and milled alloys were characterized by means of X-ray Powder Diffraction (XRD) and Scanning Electron Microscopy (SEM) both prior and after the hydrogenation process, while the hydrogen storage characteristics (P-c-T) and the kinetics were measured by using a commercial and automatically controlled Sievert-type apparatus. The hydrogenation and dehydrogenation measurements were performed at four different temperatures 150-200-250-300$^o$C and the results showed that the kinetics for both the hydrogenation and dehydrogenation process are very fast for operation temperatures 250 and 300$^o$C, but for temperatures below 200$^o$C the hydrogenation process becomes very slow and the dehydrogenation process cannot be achieved.

*Keywords: Hydrogen Storage; Metal Hydrides; Mechanical Alloying; Mg-based intermetallics;*

1. **Introduction**

Hydrogen is a potential energy carrier, and can be considered for both stationary and mobile applications [1, 2]. Furthermore, hydrogen can act as an agent allowing the energy storage resulting from renewable sources and this could be the route to overcome the production of hydrogen from fossil fuels, which causes the release of $CO_2$ [3, 4]. While it seems to be the ideal means of transport and conversion of energy for such applications, there are issues preventing the massive application of a hydrogen economy due to several safety issues and efficient storage of hydrogen [5].

Metal hydrides (MH) which can reversibly store and release hydrogen with high efficiency are considered as important materials for solving energy and environmental issues [6, 7]. In recent years a lot of research has been done on potential materials for hydrogen storage applications. Magnesium-based alloys have attracted much attention due to high hydrogen capacity, abundance and low cost [8]. It is reported that pure Mg can store up to 7.6 wt% [9-11]. Despite the high hydrogen absorption/desorption capacity, there are some serious drawbacks in the use of Mg alloys for practical on-board applications such as slow kinetics, high operation temperatures related to the high stability of the Mg–H bonds, expressed to the high enthalpy of hydride formation and high reactivity with oxygen [12, 13].

Considerable research has been conducted on magnesium alloys in order to find novel ways to synthesize new high performance materials. These studies are mainly focused on the element substitution [14], new production methods by using for example different hydrogen pressures [15, 16], preparation of composite materials through the addition of dopants in order to improve microstructure/microchemistry [17] and thermal techniques [18] in order to improve hydrogen storage characteristics.

Amongst the most efficient ways to improve hydrogen storage performance of Mg-based materials is the High-Energy Ball Milling (HEBM) [19, 20] resulting on the formation of an amorphous/nanocrystalline state of the material, synthesizing new compounds and forming composites with catalytic additions of other elements. When applying such techniques, due to the high stresses applied, the microstructure of the materials is continuously refined until a chemically and microstructure homogeneous material is obtained. Several other novel techniques have been proposed in order to synthesize high performance materials for hydrogen storage such as hybrid microwave heating [21], laser ablation techniques [22] and hydriding combustion synthesis (HCS) [23].

$Mg_2Ni$-type hydrogen storage alloys are considered to be promising negative electrode materials for nickel metal hydride (Ni-MH) batteries based on their low cost, light weight, rich mineral resources and high theoretical discharge capacity [24]. The poor kinetics and the high operating temperatures of these materials combined with the severe oxidation in alkaline solution of $Mg_2Ni$-type electrode prevent these materials for practical use. A number of studies reported that the partial substitution of Mn in a $MgNi_2$ could enhance the discharge capacity in room temperature and further to reduce the absorption process to lower temperatures [25-28].

In the current work, new ternary Mg–Mn–Ni intermetallics have been synthesised by High Energy Ball Milling and studied as potential hydrogen storage materials. Five different stoichiometries of the $Mg(Ni_xMn_{1-x})_2$ nominal composition synthesized (HEBM) under argon atmosphere at room temperature using a planetary high-energy ball miller. The microstructures were characterized by means of XRD and Scanning Electron Microscopy (SEM). The Pressure-Composition-Temperature (P-c-T) properties of these materials were measured by a commercial Sievert-type apparatus.

## 2. Experimental Procedure

Mg (99.8%, 325 mesh), Ni (99.8%, 325 mesh) (Alfa Aesar) and Mn (99.95%, 325 mesh) powders (ProChem Inc) used as the starting materials for the High-Energy Ball Milling (HEBM) process. This process was performed in high-purity Argon environment in sealed vials using commercial Fritsch "Pulverisette-7" miller and every sample was milled for 10 and 20h respectively at 350 rpm. After every hour of milling, the process was paused for 30 min in order to maintain the temperature inside the vial at low levels and after 5h of milling the process was paused again and the vial was transferred in the glove box to scratch and remove any material from the vial walls in order to ensure the homogeneity of the produced material.  The microstructure of both the as-milled and the hydrogenated samples was characterized by means of X-ray diffraction (XRD) using a BRUKER AXS D8-Advance Diffractometer.  The morphology of the samples was analyzed by Scanning Electron Microscopy (SEM) using a Zeiss NEON 40 EsB Microscope. Pressure–composition–temperature ($P$–$c$–$T$) relationships were measured by the Sieverts' method using a commercial apparatus (PCTPro 2000-HyEnergy). The purity grade of hydrogen gas used was 99.999%. Five different stoichiometries of the $Mg(Ni_xMn_{1-x})_2$ synthesized and presented in Table 1. The hydrogenation process was performed at 150-200-250-300°C at several pressures (maximum pressure 20bar) and the dehydrogenation process was performed at the same temperatures under vacuum. The activation process performed for all the samples consists of four cycles of hydrogen uptake (20 bar) and release (0.5 bar) at 300°C.

Table 1. List of all the synthesized intermetallics

|   | Sample Composition | Milling time |
|---|---|---|
| 1 | $Mg(Ni_{0.9}Mn_{0.1})_2$ | 10 – 20h |
| 2 | $Mg(Ni_{0.75}Mn_{0.25})_2$ | 10 – 20h |
| 3 | $Mg(Ni_{0.5}Mn_{0.5})_2$ | 10 – 20h |
| 4 | $Mg(Ni_{0.25}Mn_{0.75})_2$ | 10 – 20h |
| 5 | $Mg(Ni_{0.1}Mn_{0.9})_2$ | 10 – 20h |

3. Results and Discussion

3.1. Microstructure Characteristics

Figure 1 presents the XRD profiles for all the synthesised samples with the composition $Mg(Ni_xMn_{1-x})_2$ (x=0.1, 0.25, 0.5, 0.75, 0.9) milled for 10h. In addition, there is a comparison for the sample $Mg(Ni_{0.5}Mn_{0.5})_2$ for both 10h and 20h milling time (Fig1b and 1c respectively). Figure 1 indicates that the substitution of Mn instead of Ni can change the major phase $Mg_2Ni$, which is a hexagonal phase, and lead to the secondary phases Mg-Ni and Mg, where the amount of such phases is further increasing with the increase of Mn. By increasing the milling time, the peak phase $Mg_2Ni$ is broadening, which is attributed to the refined grains and stored stress that originated from the milling process.

Figure 2 shows the XRD for the sample $Mg(Ni_{0.9}Mn_{0.1})_2$ for milling times 10 and 20h respectively. The effect of the milling plays major role on the microstructure of the material. Figure 2 indicates that after 10h milling the phase $Mg_2Ni$ vanishes and the lack of that phase will play an important role during the measurement of the P-c-T properties.

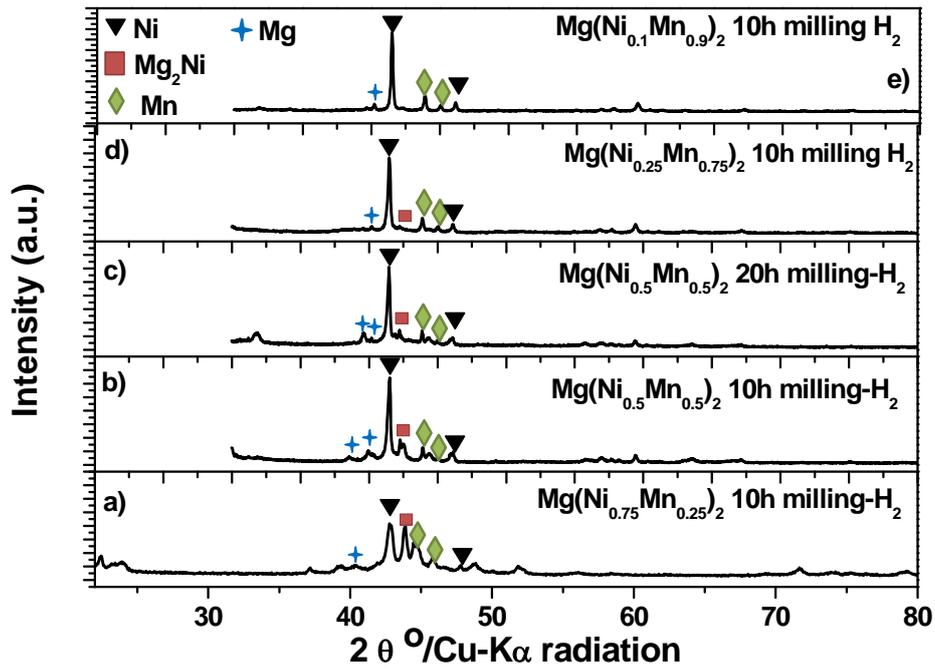

*Figure 1. XRD profiles of the alloys milled for 10h, except 1c milled for 20h*

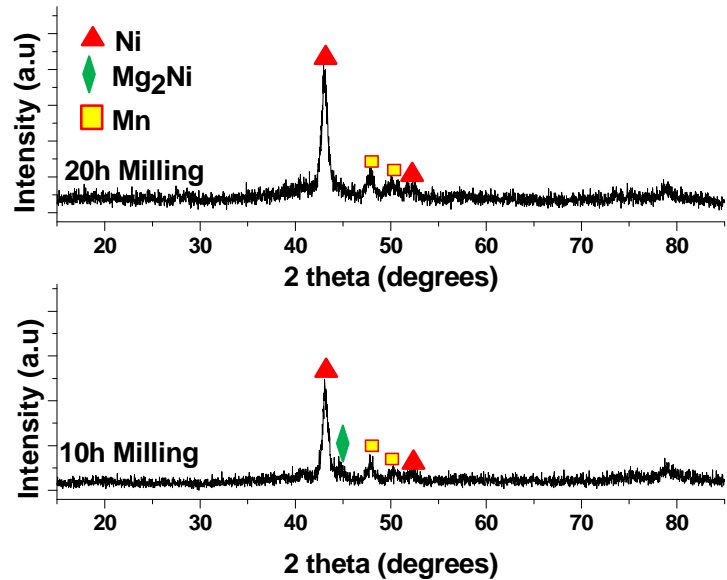

*Figure 2. XRD profiles for the sample Mg($Ni_{0.9}Mn_{0.1}$)$_2$ at 10h and 20h milling time respectively*

Finally, Figure 3 presents the XRD profile for the sample Mg($Ni_{0.75}Mn_{0.25}$)$_2$ milled for 10h before and after the hydrogenation cycling process. The presence of a new phase $Mg_2NiH_4$ is obtained indicating that the hydrogenation process was successful for the samples.

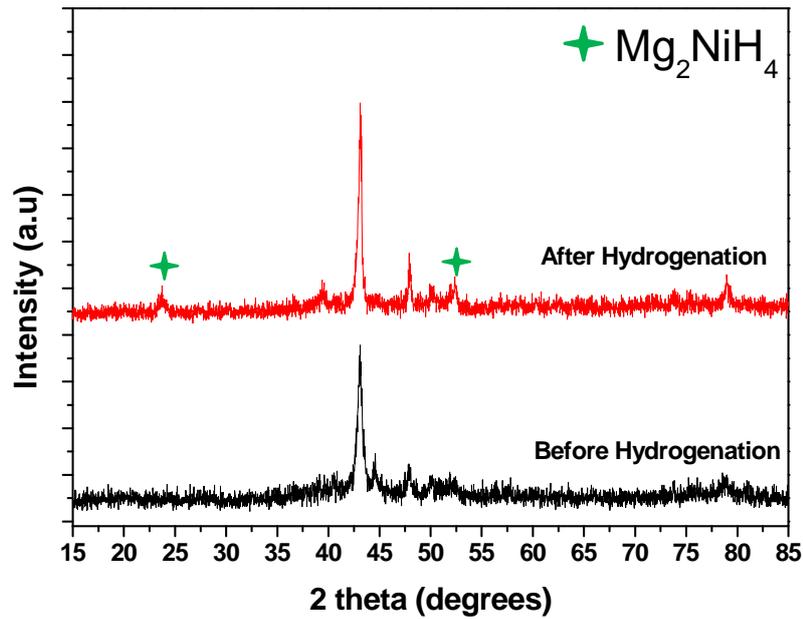

*Figure 3. XRD profiles for the sample Mg(Ni$_{0.9}$Mn$_{0.1}$)$_2$ milled for 10h before and after the hydrogenation cycling*

Figure 4 and Figure 5 shows the SEM images of the as-milled samples Mg(Ni$_{0.1}$Mn$_{0.9}$)$_2$ milled for 10 and 20h respectively. It can be seen that most of the particles (in fact, they are clusters of smaller particles) are smaller than 10 mm in size after 10h of milling. However, further milling (20 h of milling in Figure 5) does not dramatically decrease the particle size of the synthesized alloys, which indicates that a steady state equilibrium between fracturing (tending to decrease the particle size) and welding (tending to increase particle size) is attained.

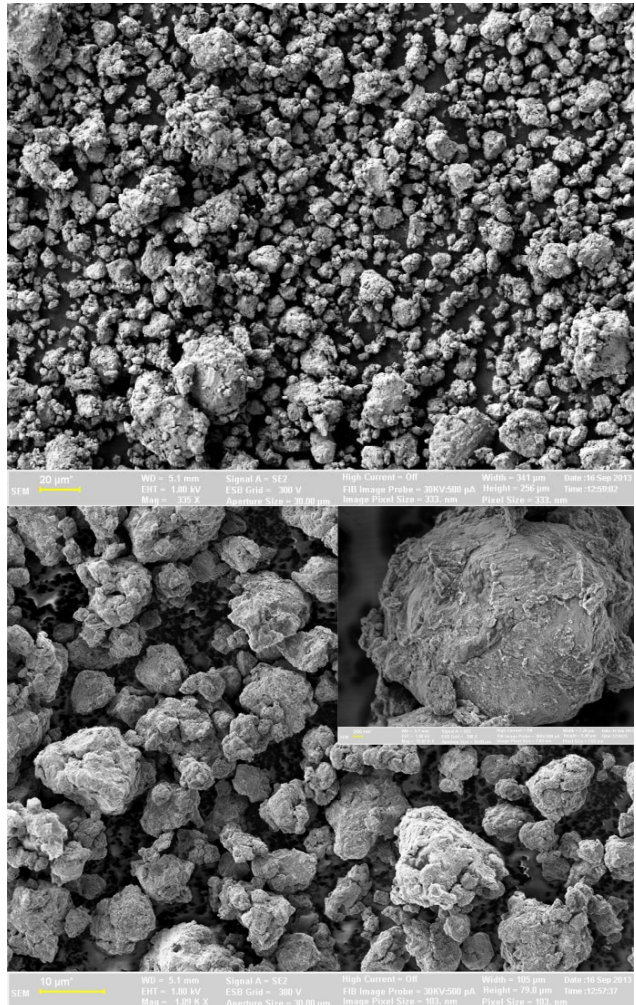

*Figure 4*. SEM micrograph of the Mg(Ni$_{0.1}$Mn$_{0.9}$)$_2$ milled for 10h

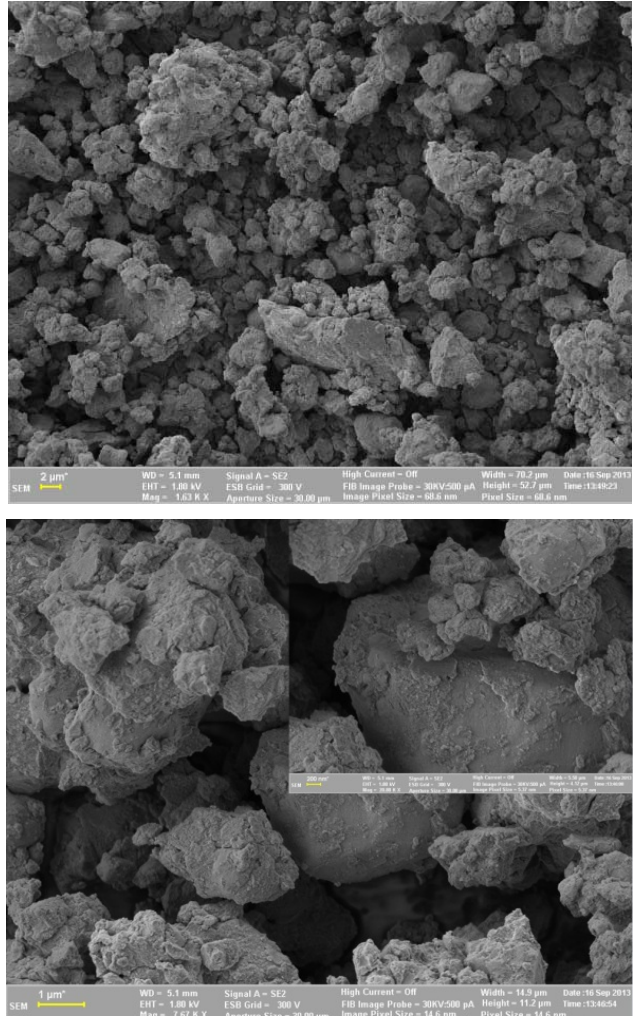

**Figure 5.** SEM micrograph of the Mg(Ni$_{0.1}$Mn$_{0.9}$)$_2$ milled for 20h

### 3.2. Sample Activation Process

The activation process consists of several cycles of hydrogen uptake at high temperature (300°C) under 20 bar of hydrogen supply pressure, followed by a hydrogen release process at the same temperature (300°C) at pressure 0.5 bar. According to the results extracted from Figure 6, for the sample Mg(Ni$_{0.1}$Mn$_{0.9}$)$_2$, after the 3$^{rd}$ hydrogenation/dehydrogenation cycle the sample is considered fully activated and the kinetics have been improved since the first two cycles. The second cycle is also fast but the hydrogen content is lower than the following cycles mainly because the dehydrogenation process of the first cycle might was not totally completed. For the dehydrogenation process, it is observed that even for the first cycle the process is very fast and the material can release all the stored hydrogen in less than 1 minute.

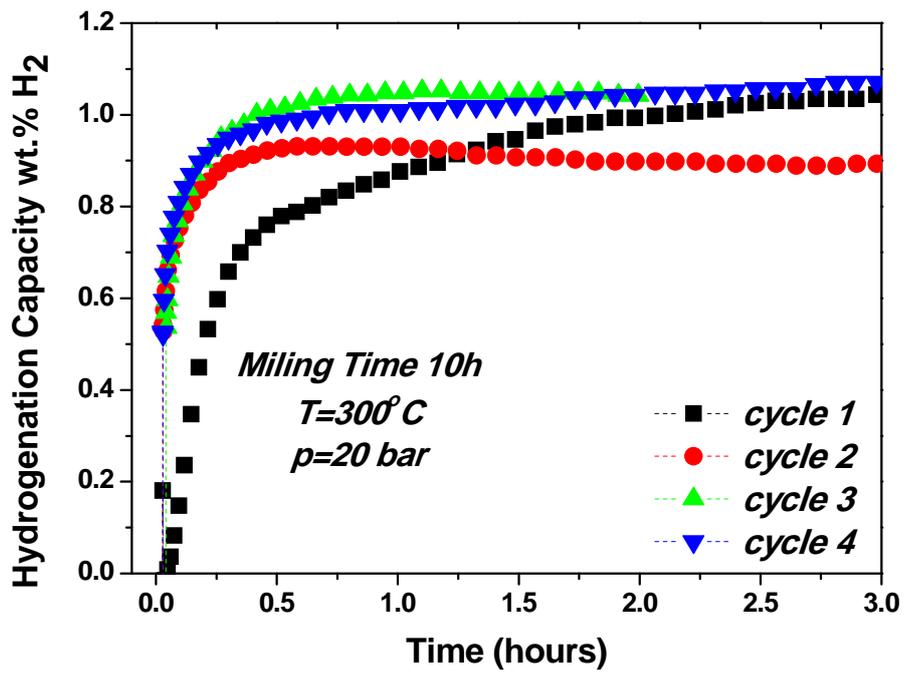

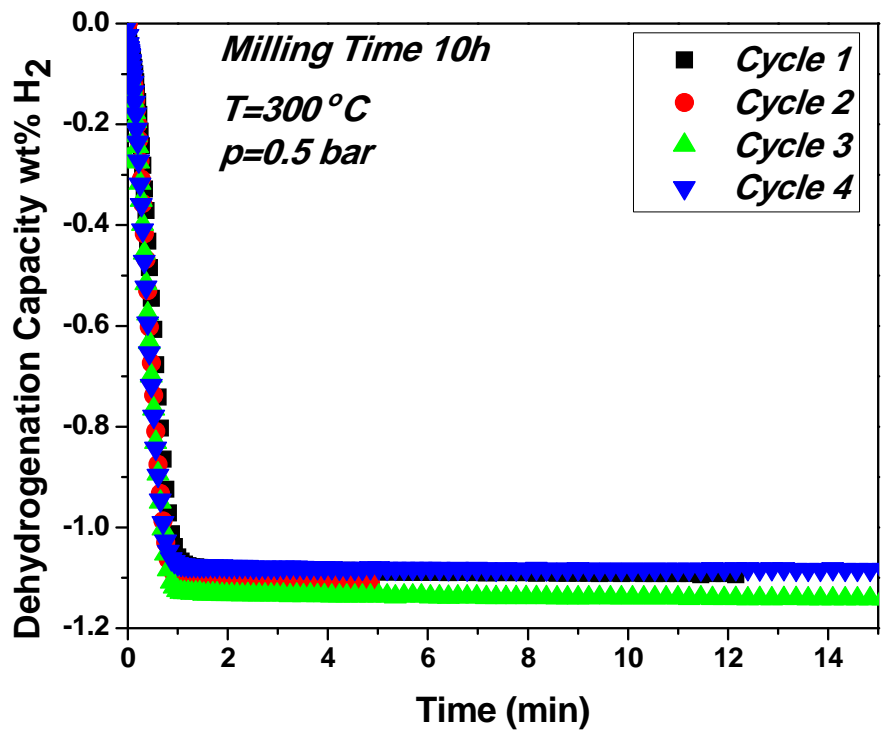

*Figure 6. Activation Cycles for the sample Mg(Ni0.1Mn0.9)2 for both the hydrogenation and dehydrogenation process*

## 3.3. P-c-T properties

Figure 7 shows the isotherms for the sample Mg(Ni$_{0.1}$Mn$_{0.9}$)$_2$ at four different temperatures 150-200-250 and 300$^{o}$C respectively for both the hydrogenation and dehydrogenation process. According to these results, the successful storage of hydrogen was achieved at all the studied temperatures, while for the release process it was impossible the release of the stored hydrogen at the lower temperatures such as 150 and 200$^{o}$C.

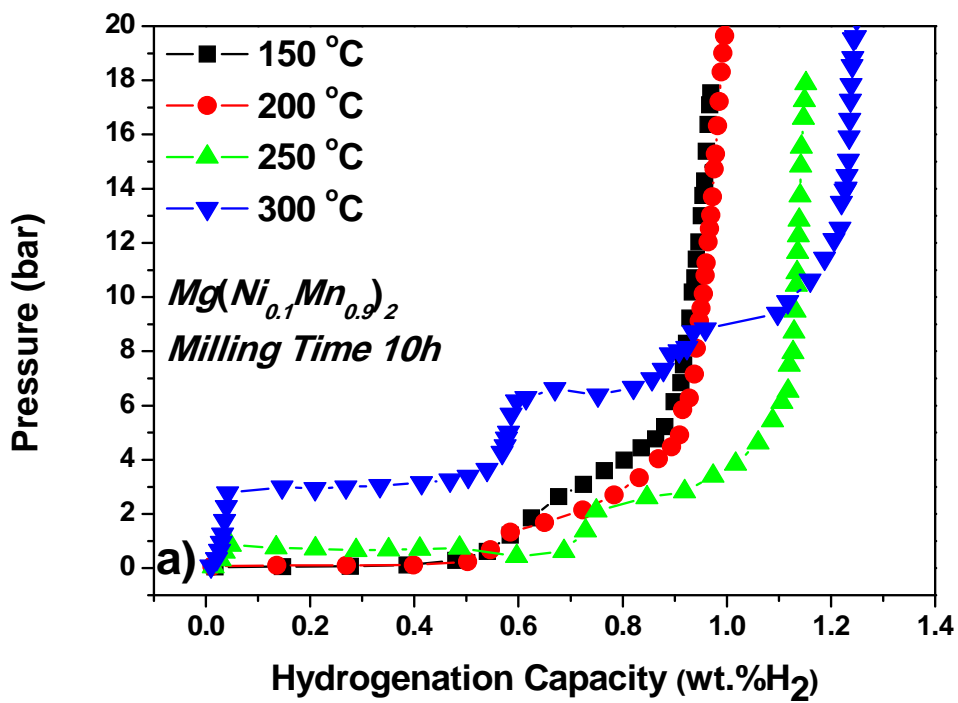

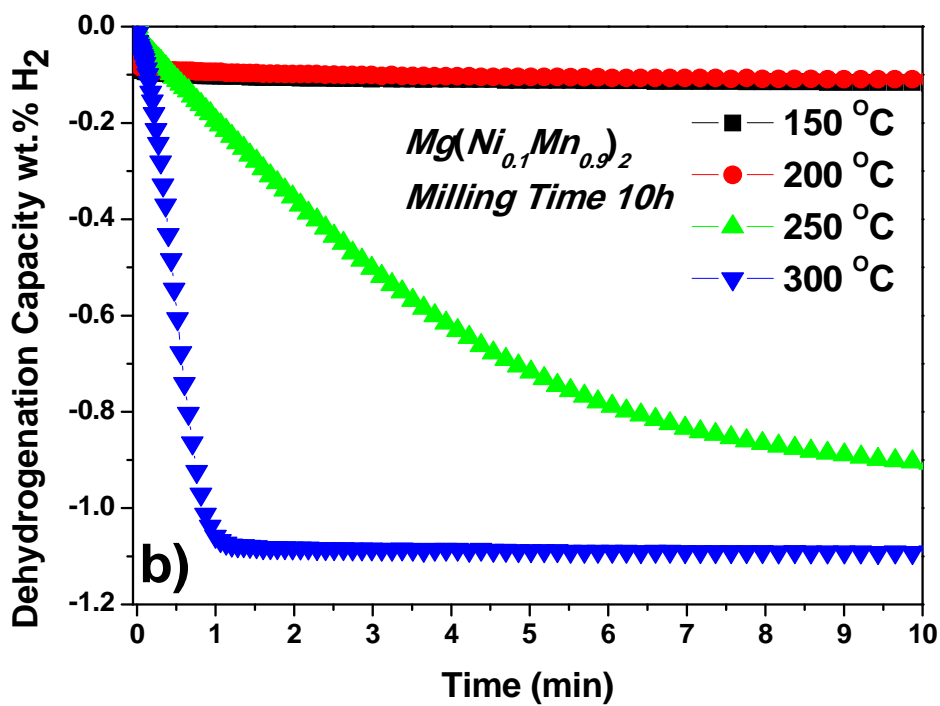

*Figure 7*. P-c-T measurements for the sample Mg(Ni$_{0.1}$Mn$_{0.9}$)$_2$ at four different temperatures for both hydrogenation and dehydrogenation process

For temperatures higher than 250°C the kinetics of the hydrogenation-dehydrogenation process improves considerably and the reaction with hydrogen is reversible. The maximum hydrogen content at 300°C is almost 1.3 wt% while at 250°C is almost 1.2 wt%. During the hydrogenation process, the main phase $Mg_2Ni$ transforms into the hydride phase as explained from the XRD analysis and also the SEM images after the hydrogenation process indicates the presence of the hydride phase (Figure 8). It is important to note that between 245 and 210°C the hydride $Mg_2Ni$-H phase transforms from a high temperature cubic structure to a low temperature monoclinic phase [29, 30]. When hydrogen is absorbed by the phase $Mg_2Ni$ beyond 0.3 H per formula unit the system undergoes a structural rearrangement to the stoichiometric complex $Mg_2Ni$-H hydride with a 32% increase in the volume.

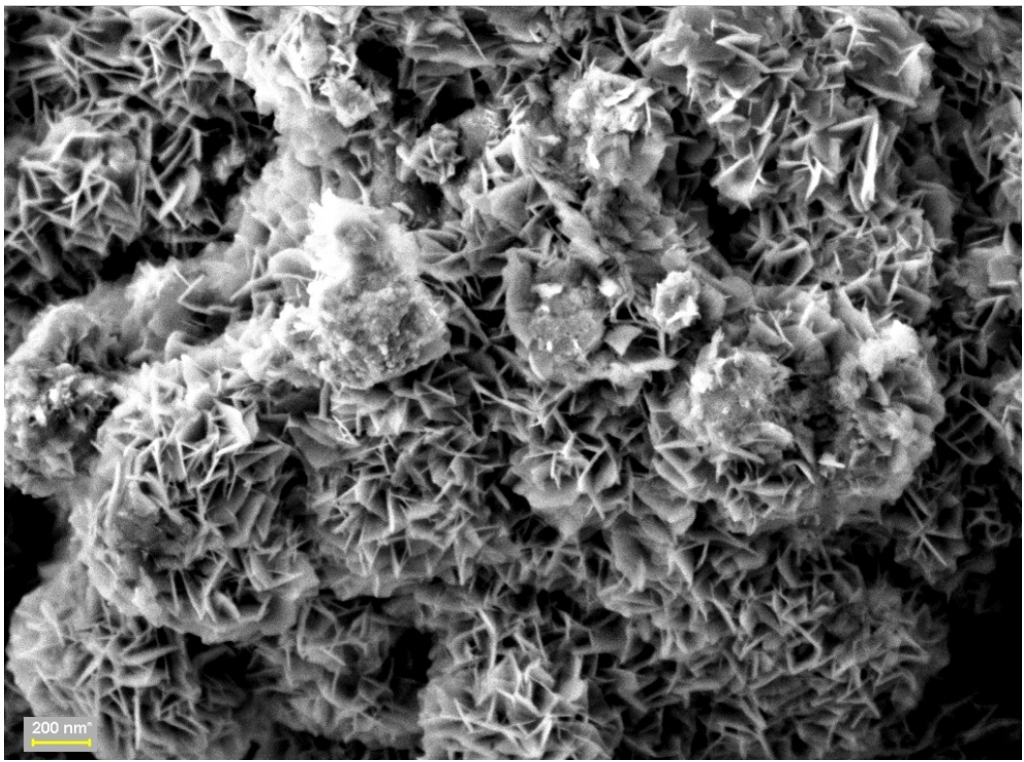

*Figure 8. SEM images after the hydrogenation process*

### 3.4. Comparison of the hydrogenation/dehydrogenation properties of the synthesized alloys.

Figure 9 presents the kinetics for all the synthesized samples for both the hydrogenation and dehydrogenation process for milling time 10h at temperature

250°C. Figure 9a shows the hydrogenation kinetics for all the synthesised intermetallics. According to the results, during the hydrogenation process, for high amount of Ni the amount of hydrogen stored is negligible and the kinetics are very slow. Especially for the sample Mg(Ni$_{0.9}$Mn$_{0.1}$)$_2$ the amount of hydrogen stored is very low and hardly can be considered as a hydrogenation process. For the sample, Mg(Ni$_{0.75}$Mn$_{0.25}$)$_2$ the amount of hydrogen stored is higher than the previous case but does not exceed 0.4 wt% and the hydrogenation process is still slow. When the amount of Mn increases more, the hydrogenation becomes faster and the amount of hydrogen stored is higher. For the sample MgMn$_2$ the maximum hydrogen content is 1.2wt% and can be achieved in during the first 10min of the process. Figure 9b presents the dehydrogenation kinetics of the synthesized intermetallics for milling time 10h at temperature 250°C and pressure 0.5bar. When operating the sample Mg(Ni$_{0.9}$Mn$_{0.1}$)$_2$ the amount of hydrogen released is very low and the process is very slow. When increasing the amount of Mn the dehydrogenation process becomes more efficient and fast and as a result, the sample MgMn$_2$ is capable of releasing all the stored amount of hydrogen in less than 5min.

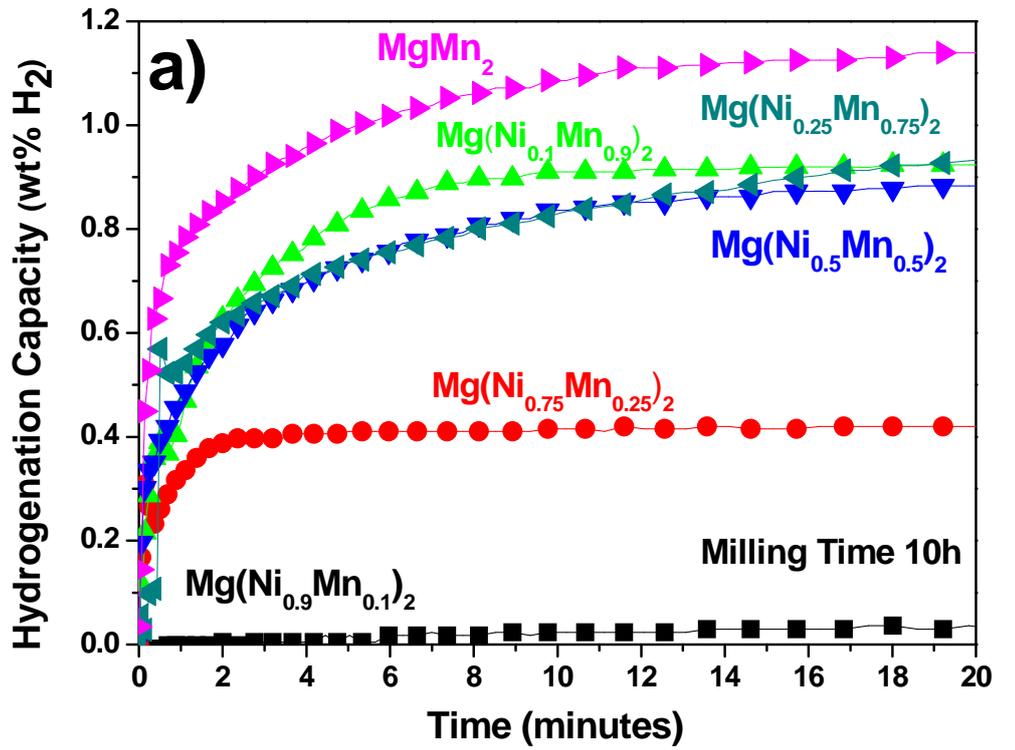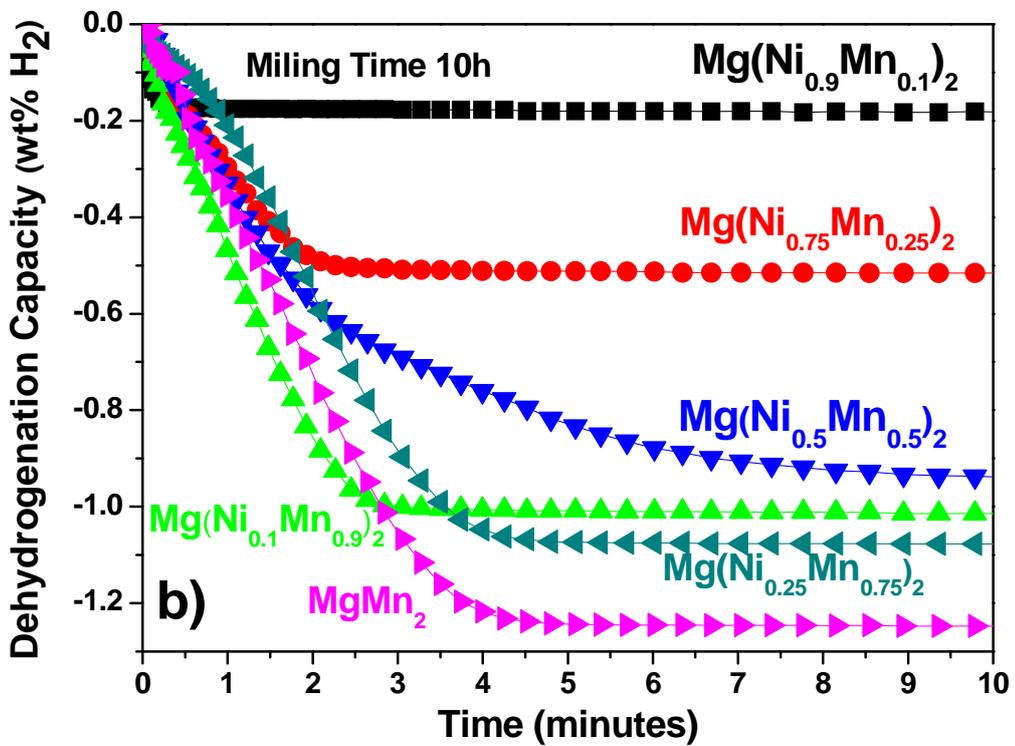

**Figure 9**. Comparison of the hydriding properties of all the synthesized alloys for milling time 10h

## 4. Conclusions

New ternary Mg-Ni-Mn alloys have been successfully synthesized by High Energy Ball Milling (HEBM) and have been studied as possible materials for efficient hydrogen storage applications. The XRD and SEM analysis proved the existence of the main $Mg_2Ni$ phase and the secondary phases Mg-Ni and Mg, where the amount of such phases is increasing with the increase of Mn. Increasing the milling time can broaden the peak for the phase $Mg_2Ni$, which is attributed to the refined grains a stored stress that originated from the milling. According to the SEM images it was concluded that further milling beyond 10h does not dramatically decrease the particle size of the synthesized alloys, which indicates that a steady state equilibrium between fracturing (tending to decrease the particle size) and welding is attained. The hydrogenation/dehydrogenation kinetics and the amount of hydrogen that can be stored/released depends hardly on the amount of Mn. For a small amount of Mn the kinetics for both the hydrogenation and dehydrogenation are very slow and the amount of hydrogen stored/released is negligible. When increasing the amount of Mn the kinetics are faster and the materials are able to store/release the full amount of hydrogen at operation temperature $250^oC$.

**Acknowledgements**

This work has been supported from the Greek Ministry of Scientific Research and Technology, Greece-Romania bilateral project, code: **11ROM 1_3_ET 30**